# EXPRESSIVENESS OF A PROVENANCE-ENABLED AUTHORIZATION LOGIC

Jinwei Hu

School of Computer Science and Technology,
Huazhong University of Science and Technology, Wuhan, China
jwhu@hust.edu.cn

## ABSTRACT

*In distributed environments, access control decisions depend on statements of multiple agents rather than only one central trusted party. However, existing policy languages put few emphasis on authorization provenances. The capability of managing these provenances is important and useful in various security areas such as computer auditing and authorization recycling. Based on our previously proposed logic, we present several case studies of this logic. By doing this, we show its expressiveness and usefulness in security arena.*

## 1. INTRODUCTION

The popularity of Internet has promoted wide applications of distributed systems such as peer-to-peer systems and grid. These systems feature dynamics, provisionality, lack of complete information, and decentralization, consequently they introduce more complex authorization scenarios. It is widely agreed that a declarative and reasonably expressive language with an unambiguous semantics would be a key to distributed access control. Based on this observation, major research efforts have applied logics into the design of authorization languages. The set of policies written in such a language is regarded as a policy base. When a principal requests an access to resources, the request is translated to a query of the policy base. The access is granted if the answer to the query is positive and denied otherwise.

One main line of preceding languages depends on Datalog and its variants, such as DL, RT, Binder, and SecPAL [11, 12, 5, 3]. Generally speaking, these languages achieve a balance between the expressiveness and computational tractability. However, they hardly capture some important aspects of distributed authorizations. Consider a delegation "A says B cansay f" in SecPAL. The delegable fact f can only be an atom here. No rule such as $f_1 \Rightarrow f_2$ is allowed to delegate. Not only may an agent trust others in the truth of facts, it may also depend on others' judgement about the relation between facts. For example, a bookshop assistant may rely on her manager about whether a discount could be given to students; that is "Assistant says Manager cansay $discount \Rightarrow student$". In addition, their reasoning ability is confined to administrators' positive knowledge about facts. Let F denote the set of facts in a SecPAL policy base. All authorizations are conditional on the administrator's knowledge about facts in F, i.e., "Admin says f", either derived or directly phrased. However, due to the lack of complete information in distributed environments, some authorizations have to rely on administrators' partial knowledge or their knowledge about others' knowledge. For instance, the bookshop assistant knows that a book may be sold only if she knows that the manager knows that the price is fair, but not just if she knows that the price is fair; again, the bookshop assistant knows that a student may borrow a book for a two-day preview if she considers it is possible that the student will return the book.

Another mainstream is to interpret policies using the Kripke structures [2, 6]. In previous works, belief has been used implicitly or explicitly to facilitate distributed access control. Most logics for distributed authorizations are centered around "A says f". Intuitively, the meaning of "A says f" is that A supports f, i.e., A believes that f is true. Also, in [1] Abadi stated that "In general, the proof may exploit relations between A and B and other facts known to the reference monitor". We argue that partial and nested knowledge besides in addition to pure facts are crucial to make correct access control decisions. There is already a potential trend to make beliefs explicit. Gurevich and Neeman remarked that knowledge should be made explicit due to its importance in authorization languages [7]. Besides, in [6] Garg and Abadi showed that Kripke semantics leads to several advantages such as comparisons of access control logics.

Since delegations provide a flexible way to evaluate the trustworthiness of a statement in distributed environments, most current policy languages support delegating capabilities. Briefly, delegations enable agent A to speak on behalf of agent B with respect to a certain statement, say $\varphi$. Delegations can also form a delegation chain with certain length. Generally speaking, we can assume the agent at the starting point of a delegation chain is in charge of the requested resources. We refer to the set of agents who issue the delegations appearing in a delegation chain as a provenance because they actually describe from where beliefs are concluded. We observe that existing languages do not distinguish between the same conclusions but with different provenances. Consider the policy base PB {"Alice says Bob cansay deletefile", "Bob says deletefile", "Alice says Charlie cansay deletefile", "Charlie says deletefile"} written in SecPAL. Then from this PB, one can conclude "Alice $says_\infty$ deletefile". But it is not clear whether it is with the help of Bob or of Charlie that PB comes to this conclusion.

There are several reasons why it is crucial to make clear conclusions' provenances. First, it is possible to enforce fine-grained constraints on provenance. For example, constraints forbidding certain provenances appear difficult to be enforced in previous approaches. Second, one may trust more than one agent in some facts and to different degrees, especially when policies are specified by multi-authors. And how much trust would suffice depends on the requested accesses. For instance, the bookshop may delegate the fact that "David is a student" to both the registrar and the professor Emma. Obviously, the registrar is at least as trustworthy as Emma in this respect. Emma's statement that "David is a student" is enough for the bookshop to give a student discount to David, whereas the registrar's testimony is needed when David wants to borrow books. Third, by indicating precisely who executes the delegated authority, we provide a more useful log if proofs of authorization decisions are included in the log [16], therefore it would be easier to work out who is responsible for which statements and derivations.

In our previous works [8, 9], we proposed an access control logic DBT based on the classical KD45 belief framework. DBT extends the BT logic [14] by introducing a new modal operator $D_i$ for each agent i into the underlying distributed authorizations. $D_i\varphi$ is designed to express the provenance of $\varphi$. Thus, DBT integrates the belief, trust, and provenance within a unified logical framework. In this paper, we present several case studies of this logic, including examples for subjective attributes and incomplete information. By doing this, we show its expressiveness and usefulness in security arena.

The rest of the paper is organized as follows. The background is introduced in Section 2. The motivation of authorization provenance is illustrated in a case study in Section 3, followed by a case study of subjective attributes in Section 4 and discussions of policies with incomplete information in Section 5. Finally, we conclude in Section 6.

International Journal of Managing Information Technology (IJMIT), Vol.2, No.2, May 2010

## 2. BACKGROUND: THE LOGIC DBT

Consider a finite set of agents AG = {1, · · · ,N}. We have three types of modal operators for each agent i: $B_i$, $D_i$, and $T_j^i$. $B_i\varphi$ means that agent i believes $\varphi$ or that i says $\varphi$; and $T_j^i\varphi$ reads that agent i trusts agent j on or that i delegates $\varphi$ to j. $D_i\varphi$ means that "due to agent i, $\varphi$ holds" or that i causes that $\varphi$ holds. A subset AE of AG is called an agent expression. Given an AE AG, we also define an operator $D_{AE}$ based on $D_i$ for each i AE. $D_{AE}\varphi$ means that the set AE of agents together cause $\varphi$. Let Prop be a set of primitive propositions. The set of well-formed formulas is inductively defined as follows:

$$\varphi ::= p \mid \neg\varphi \mid \varphi \wedge \varphi \mid \varphi \Rightarrow \varphi \mid B_i\varphi \mid D_i\varphi \mid D_{AE}\varphi \mid T_j^i\varphi$$

Readers are referred to [8, 9] for details.

## 3. THE USAGE OF OPERATOR $D_{AE}$

The most distinguished feature of DBT is the introduction of the operator $D_{AE}$. We now show the intuition of $D_{AE}$ through examples.

Suppose that a company manages two resources: computers and printers. This company lets two security administrators, SA1 and SA2, specify policies for accessing computers and printers, respectively. SA1 specifies that a clerk acknowledged by the human resources, HR, may be allowed to access printers. But SA2 thinks Manager's attestation is enough for a clerk to use computers. Putting together, the company's policy base may be as follows.

$$T_{HR}^{Company}\, \mathsf{clerk(Alice)} \qquad (1)$$

$$B_{Company}\, \mathsf{clerk(Alice)} \Rightarrow B_{Company}\, \mathsf{access}(printers) \qquad (2)$$

$$T_{Manager}^{Company}\, \mathsf{clerk(Alice)} \qquad (3)$$

$$B_{Company}\, \mathsf{clerk(Alice)} \Rightarrow B_{Company}\, \mathsf{access}(computers) \qquad (4)$$

Suppose that Manager issues to Alice a credential (5) which says Manager believes Alice is a clerk. If interpreting by some existing logics such as BT in [14], from (3) and (5), one can derive that Company believes that Alice is a clerk. As a result, Alice may log onto a computer because of (4); but from (2), Alice may also use printers. The latter authorization departs from the policies in that SA1's intention on accesses to printers is not enforced; it also violates the principle of least of privilege.

$$B_{Manager}\, \mathsf{clerk(Alice)} \qquad (5)$$

A possible solution to handle this difficulty is to rename the attribute "clerk(Alice)" in (1) and (2) as "HR-clerk(Alice)", and that in (3) and (4) as "Manager-clerk(Alice)", respectively. Then Manager only says that he/she believes Manager-clerk(Alice), which derives that Company believes Manager-clerk(Alice) together with the renamed (3), but not that Company believes HR-clerk(Alice) with the renamed (1). But this solution is ad hoc and highly dependent on implementation, and complicates the analysis of policy bases.

Alternatively, one may attempt to assemble Company's policy base instead as follows:

$$B_{HR}\text{clerk(Alice)} \Rightarrow B_{Company}\text{access}(printers) \qquad (6)$$

$$B_{Manager}\text{clerk(Alice)} \Rightarrow B_{Company}\text{access}(computers) \qquad (7)$$

This policy base may derive improper conclusions, however. We can assume that the statement (6) is from Company; that is Company bases its belief on others' beliefs. However, from (6) it follows that

$$\neg B_{Company}\text{access}(printers) \Rightarrow \neg B_{HR}\text{clerk(Alice)}. \qquad (8)$$

(8) is HR's assertion. As a result, Company can derive HR's belief from its own belief. If, for some reasons, Company forbids Alice from using printers by issuing $B_{Company} \neg access(printers)$. Then, one may deduce $\neg B_{HR} clerk(Alice)$; that is, HR does not believe clerk(Alice). Agents should not derive others' belief simply from their own belief. Interestingly, some recent works on policy languages [3, 7] avoid this form of policies as well.

Our solution is to replace (2) and (4) with the following rules.

$$D_{HR}B_{Company}\text{clerk(Alice)} \Rightarrow B_{Company}\text{access}(printers) \qquad (9)$$

$$D_{Manager}B_{Company}\text{clerk(Alice)} \Rightarrow B_{Company}\text{access}(computers) \qquad (10)$$

From (5), it follows that Manager causes Company to believe that clerk(Alice), which, together with (10), implies that Company believes that Alice can access computers. However, with (5), one can not derive that HR causes Company to believe that clerk(Alice).

The essential reason of this violation is that there are multiple trust sources inherently for some attributes. For example, the local party may trust both school's Registrar and Professor to say student(Bob). This is not exclusive to multi-author policy bases, though. On the other hand, to achieve availability and resilience, the local party may intentionally delegate the identification of attributes to more than one party.

Despite agents may be delegated with the identification of the same attribute, they differ in the extent to which Local may trust them about the attribute. For example, credentials about an attribute student(Bob) from both school's Registrar and Professor may be trusted by Local, but to different extents and for different purposes. Owing to the difference of importance and sensitiveness among resources, the local party may consider credentials from different trustees as sufficient for accesses to resources to be granted, even though these credentials support the same attribute. However, when putting together in one policy base, it is likely that these policies affect each other. The introduction of "due to" can alleviate this influence, and enable the local party to safely base authorizations on the same set of attributes.

## 4. SUBJECTIVE ATTRIBUTES

Subjective attributes are specially demanded in some distributed systems, such as peer-to-peer networks and electronic commerce systems. In these systems, agents may express opinions about other agents' behaviors. Looking from different perspectives, agents may view one happening in totally different ways, or even in conflicted ways. For example, Alice may regard Cathy a good peer because of uploading some files, whereas Bob could think of this behavior offending. However, it is not something right or wrong, but simply subjective. Thus, authorization logic should be able to accommodate this divergence so that policies about subjective attributes are amendable to formal analysis. For better explanation, we illustrate this by an example.

To support subjective attributes, policy bases should satisfy two requirements. On one hand, trustees should have discretion to judge the truth and falsity of an attribute. For example, if Alice puts trust on Bob about whether or not goodPeer(David), Bob should be free to say that he believes goodPeer(David) or the opposite. On the other hand, Local should respect the divergence of viewpoints among trustees. Assuming that Alice also trusts Cathy about goodPeer(David), if Bob and Cathy have opposite opinions about goodPeer(David), Alice should treat this case as normal but not conflicted.

Consider a group of four peers in a peer-to-peer network, Alice, Bob, Cathy, and David. Alice has a policy base which controls the accesses to her shared directories. Alice trusts both Bob and Cathy to tell whether or not David is a good peer. When both of them consider David as good, Alice permits accesses to dir; but if only one of them thinks so, only access to subdir1 or subdir2 is allowed.

Let $\varphi = goodPeer(David)$ and let $\varphi_0 = canAccess(David, dir)$, $\varphi_1 = canAccess(David, subdir1)$, and $\varphi_2 = canAccess(David, subdir2)$. Using DBT, we may specify policies as follows:

$$T_{Bob}^{Alice} \varphi \qquad T_{Bob}^{Alice} \neg \varphi \qquad (11)$$

$$T_{Cathy}^{Alice} \varphi \qquad T_{Cathy}^{Alice} \neg \varphi \qquad (12)$$

$$D_{Bob} B_{Alice} \varphi \land D_{Cathy} B_{Alice} \varphi \Rightarrow B_{Alice} \phi_0 \qquad (13)$$

$$D_{Bob} B_{Alice} \varphi \land D_{Cathy} B_{Alice} \neg \varphi \Rightarrow B_{Alice} \phi_1 \qquad (14)$$

$$D_{Bob} B_{Alice} \neg \varphi \land D_{Cathy} B_{Alice} \varphi \Rightarrow B_{Alice} \phi_2 \qquad (15)$$

Previous policy languages rarely support subjective attributes. To meet the first requirement, Alice's policy base should contain both $T_{Bob}^{Alice} goodPeer(David)$ and $T_{Bob}^{Alice} \neg goodPeer(David)$. However, most Datalog-based policy languages trade off the ability to delegate negative attributes for other advantages like computational tractability one one hand; on the other hand, some access control logics forbid delegating both $goodPeer(David)$ and $\neg goodPeer(David)$ to the same agent [6]. Without this capability, agents are not able to express their subjective judgements. More importantly, supposing that Alice makes the delegations in (11) and (12), and that both $B_{Bob} goodPeer(David)$ and $B_{Cathy} \neg goodPeer(David)$ reside in Alice's policy base, existing access control logics would deduce conflicts. In spite of the possibility of agents lying, it is natural for agents to have different (or even opposite) viewpoints for subjective attributes.

Since Local lacks of complete information, it depends on other agents' statements. Here, Alice believes $goodPeer(David)$ because of Bob, whereas Cathy also causes Alice to believe $\neg goodPeer(David)$. The point is that Alice has two ways to collect information. When objective attributes are involved, it is reasonable to conclude policy bases conflicted in this case and employ some conflict resolving mechanisms. However, as far as subjective attributes are concerned, it is more acceptable to let the local party put different weights on how information is aggregated.

In the DBT representation above, the semantics allows Alice to delegate both $goodPeer(David)$ and $\neg goodPeer(David)$ to Bob. We use the operator $D_i$ to denote the subjective judgements as in (13), (14) and (15). For example, when both $B_{Bob}goodPeer(David)$ and $B_{Cathy}\neg goodPeer(David)$ are in Alice's policy base, (11) and (12) derive $D_{Bob}B_{Alice}goodPeer(David)$ and $D_{Cathy}B_{Alice}\neg goodPeer(David)$, respectively. Thus, it follows from (14) that Alice believes that David can access subdir1.

## 5. POLICIES WITH INCOMPLETE INFORMATION

There are some attributes whose member would be denied access explicitly because their accessing may incur great loss to Local. For example, an airport, denoted as Local, would forbid terrorists to board. We refer to requesters bearing these attributes as adversaries. The problem is that, an adversary would withhold some credentials to evade explicit denials, and adversary authorities would not inform Local of adversaries. No terrorist, of course, would reveal this attribute and terrorist organizations would not inform the airport.

The airport depends on its Scrutiny Unit (SU) to identify potential terrorists. With existing policy languages, one tempting solution is to include $T_{SU}^{Local}terrorist(X)$ or (and) $T_{SU}^{Local}\neg terrorist(X)$ in the policy base. However, we argue that it is unreasonable for Scrutiny Unit to be able to say a passenger is or not. Because, in most cases, Scrutiny Unit may only discover some clues showing that a passenger is possibly a terrorist. That is, Scrutiny Unit generally only suspects someone. Neither $B_{SU}terrorist(X)$ nor $B_{SU}\neg terrorist(X)$ captures the meaning of "suspect".

In DBT, the airport's policy bases may consist of the following formulas.

$$T_{SU}^{LOCAL}\neg B_{SU}\neg terrorist(X) \quad (16)$$
$$T_{SU}^{LOCAL}\neg B_{SU}terrorist(X) \quad (17)$$
$$D_{SU}B_{LOCAL}\neg B_{SU}\neg terrorist(X) \Rightarrow B_{LOCAL}\neg Permit(X, board) \quad (18)$$
$$D_{SU}B_{LOCAL}\neg B_{SU}terrorist(X) \Rightarrow B_{LOCAL}Permit(X, board) \quad (19)$$
$$\neg B_{SU}terrorist(X) \quad (20)$$
$$\neg B_{SU}\neg terrorist(X) \quad (21)$$

From (16) and (17), Local delegates to Scrutiny Unit whether or not Local should consider a passenger as a potential terrorist; and by (18) and (19), Local explicitly denies or permits passengers to board, respectively, according to Scrutiny Unit's statements. If a passenger obtains a certificate from Scrutiny Unit like (20), the airport allows him/her to board; But if anyone get a certificate (21), he/she would be denied to board, whether or not he/she submit the certificate.

Obviously, there may be both false positives and false negatives; Scrutiny Unit may issue (21) to a non-terrorist and (20) to a real terrorist. Since no complete information is available, it is unavoidable. But the rate depends on the trusted agent and the requested resources. Besides, as for false negative, other techniques, such as audit and intrusion detection alarm, may be deployed, which is beyond the scope of this work.

## 6. CONCLUDING REMARKS

We illustrated DBT is able to capture provenance-aware scenarios. This expressiveness is the original motivation of DBT and is the most important feature. Existing access control logics put few emphases on authorization provenances. As mentioned in Section 1, existing logics can be grouped into the modal logic based and the Datalog based frameworks. For the first group, since they are interpreted in the same framework as DBT, it seems feasible to extend these logics to express provenance or to build a new logic on them. Take the logic ABLP [2] for instance. ABLP works around two operators: "says" and "speakfor". Since formulas constructed using these two operators are interpreted by Kripke structures, we may define operators for provenances and impose some reasonable relations among modalities. The other collection based on Datalog includes Delegation Logic (DL) [11], SecPAL [3], and RT [13], ect.. This group features in tradeoff between reasonable expressiveness and tractability. Nevertheless, none of these policy languages focuses on what the operator $D_i$ is designed to capture. It also appears difficult to incorporate the notion of authorization provenances into these logics. Because specific translation rules between these logics and Datalog may have to be redesigned; another requirement on these rules is that the resulted semantics should bring about rational connections among agents' statements, authorizations, and provenances. Still, a notion of proof tree is used in literature. SD3 [10] produces a proof tree along with the answer to each query to see if the proof is correct. $RT_0$ [13] forms delegation chains for a policy base, but its focus is on how to store and retrieve credentials in a distribute way. Neither of them can answer if a conclusion with a specific provenance holds.

As future work, we are planning to integrate authorization provenances into auditing and recycling, and invest how well quantitatively provenances benefit these functions through experiment evaluation.

**Authors**

Jinwei Hu is a PhD student in Huazhong University of Science and Technology.